\documentclass[preprint2]{aastex}
\usepackage{epsfig}
\usepackage{graphicx}
\usepackage{color}
\usepackage{enumerate,amsmath}
\newcommand{\kms}{{\rm km\,s^{-1}}}
\newcommand{\beq}{\begin{equation}}
\newcommand{\eeq}{\end{equation}}
\newcommand{\ba}{\begin{eqnarray}}
\newcommand{\ea}{\end{eqnarray}}

\def\msun{M_\odot}
\def\rsun{R_\odot}
\def\mb{M_\bullet}
\def\mbm{M_{\bullet,m}}
\def\mbs{M_{\bullet,s}}
\def\jlc{J_{\rm lc}}

\def\thelc{\theta_{\rm lc}}
\def\jtd{J_{\rm td}}
\def\jmb{J_{\rm mb}}

\def\rmax{r_{\rm max}}
\def\rtr{{r_{\rm tr}}}

\def\peryr{{\rm yr^{-1}}}
\def\E{{\cal E}}
\def\rcri{r_{\rm cri}}

\shortauthors{F.K. Liu and Xian Chen}
\shorttitle{{Enhanced} off-center tidal disruptions in merging galaxies}

\begin{document}
\title{{Enhanced} off-center stellar tidal disruptions by 
    supermassive black holes in merging galaxies} 
 
\author{F.K. Liu\altaffilmark{1} and Xian Chen\altaffilmark{2,3}}

\altaffiltext{1}{Department of Astronomy, Peking University, 100871
  Beijing, China; {\it fkliu@pku.edu.cn}} 
\altaffiltext{2}{Kavli Institute for Astronomy and Astrophysics,
  Peking University, 100871 Beijing, China; {\it chenxian@pku.edu.cn}} 
\altaffiltext{3}{Max-Planck Institute for Gravitational Physics
  (Albert Einstein Institute), 14476 Golm, Germany} 

\begin{abstract}

Off-center {stellar} tidal {disruption} flares have been
  suggested to be {a} powerful 
probe of recoiling {supermassive black holes (SMBHs)} 
out of galactic centers 
due to anisotropic gravitational wave radiations. However, off-center
tidal flares can also be produced by SMBHs in merging galaxies. In 
this paper, we computed the tidal flare rates by dual SMBHs in
  {two} merging
galaxies before the SMBHs become {self-gravitationally}
bounded. We employ an analytical model to calculate the tidal
loss-cone feeding rates for both SMBHs, taking into account two-body
  relaxation of stars, tidal 
perturbations by the companion galaxy, and chaotic stellar orbits in 
triaxial gravitational potential. We show that {for typical SMBHs with masses $10^7~\msun$, the loss-cone feeding rates are enhanced by mergers} up to $\Gamma \sim 10^{-2} \, {\rm yr^{-1}}$, about
two order of magnitude higher than those by {single} SMBHs in isolated
galaxies and about four orders of magnitude higher than
those by recoiling SMBHs. The enhancements are mainly due to
tidal perturbations by the companion {galaxy}. We suggest that
off-center tidal flares are overwhelmed by those from merging galaxies, 
{making} the identification of recoiling SMBHs challenging. Based on
the calculated rates, we estimate the relative contributions of 
tidal {flare events}  by single, binary, and
  {dual}  SMBH systems during cosmic time. Our
  {calculations show} that the 
  off-center tidal disruption flares by un-bound SMBHs in merging
  galaxies contribute a fraction comparable to that by 
single SMBHs {in} isolated {galaxies}. {We conclude
  that} {off-center
  tidal disruptions are powerful tracers of the merging history
  of galaxies and SMBHs.} 

\end{abstract}


\section{Introduction}\label{sec:intro}

In the $\Lambda$-cold dark matter cosmology, both dark matter halos and
  galaxies form due to frequent mergers. In this paradigm, 
hierarchical galaxy mergers would incorporate multiple supermassive
  black holes (SMBHs) into a galaxy \citep{vol03}. 
When two SMBHs, initially embedded in the two cores of the merging
  galaxies, sink to the common center of the system due to dynamical
  friction and become gravitationally bound, a supermassive black hole
  binary (SMBHB) would form \citep{beg80}. During the interaction
  between the SMBHB and the stellar and gaseous environments, if the
  two SMBHs could successfully evolve to a separation of hundreds of
  Schwarzschild radius, then gravitational wave (GW) radiation could
  lead to the coalescence of the SMBHs within a Hubble time, and the
  asymmetry of GW radiation is predicted to impart a recoiling
  velocity on the post-merger SMBH \citep{hughes09,cen10}. Detection
  of the GW radiation from coalescing SMBHB would be a vital test of
  the theory of general relativity (GR), and is the major goal of the
  ongoing Pulsar Timing Array (PTA) project and {any future
  space-based GW mission.} 

Despite many efforts to detect GW radiation from coalescing SMBHBs,
theoretical studies found large uncertainties for the dynamical
evolution of SMBHB in normal galaxies: in the absence of gas and
efficient stellar relaxation, the evolution of SMBHB would stall at
sub-parsec (pc) scale and not enter the GW radiation regime
\citep{mer08,col09}, while recent $N$-body simulations suggest that
efficient repopulation of stars to the galaxy core may be norm in
real mergers \citep{preto11,khan11}. Observationally, it is difficult
to test the dynamical evolution of SMBHB in stellar systems, because
of the lack of electromagnetic (EM) radiation from the vicinity of the
dormant SMBHs. Recently, ``tidal flares'', the EM outbursts produced
due to tidal disruption of stellar objects by SMBHs, have been
identified as powerful probes of the mass and spin of the otherwise
dormant SMBHs
\citep{rees88,kom02,don02,van11,gez09,blo11,bur11,cen11}. The 
flaring rate for a single SMBH in an isolated galaxy is
  estimated to be $10^{-5}$ to $10^{-4}~\peryr$ 
\citep{magorrian99,syer99,wang04,brockamp11}. 

It is predicted that the formation and evolution of bound
SMBHB at galaxy center would significantly change the event rate and
affect the light 
curves of tidal flares. Shortly after the formation of SMBHB, the
three-body interaction between the binary and a bound stellar cusp
will enhance the flaring rate to as high as $1~\peryr$
\citep{ivanov05,chen09,chen11,wegg11}. After the stellar cusp is
disrupted, mainly due to  slingshot ejection, if stellar relaxation is
inefficient in galaxy center, the flaring rate will become one order
of magnitude lower than that in the single black hole system
\citep{chen08}. When the SMBHB enters the GW-radiation regime, the
interruption and recurrence of tidal-flare light curve by the
perturbing secondary 
black hole occur on an observable timescale \citep{liu09}, and the
stars resonantly trapped by the inspiralling SMBHB may produce a tidal
flare around the coalescence of the binary \citep{sch10,seto11}. After
the coalescence of the binary, the launch of a recoiling SMBH
might be also accompanied by a brief burst of tidal flares. 
As the recoiling SMBH travels
outside the galaxy core, tidal disruption of the stars gravitationally
bound to the hole may produce a flare apparently displaced from the
galaxy center \citep{komossa08,mer09,stone11b,li12}. Because of the
many differences between the flaring rates in single and binary SMBH
systems, it is suggested that tidal flares can be utilized to
constrain the fraction and dynamical evolution of SMBHBs in galaxy
centers \citep{chen08}. 

Off-nuclear tidal disruption flares and ultra compact star
clusters with
  peculiar properties are suggested to be the key features of
  gravitational recoiling SMBHs in galaxies. However, off-center tidal
  disruption flares can also be produced by SMBHs 
  in merging galaxies, and star clusters with the proposed
  peculiar characters 
may also form by tidal truncation of secondary galaxies in minor
  mergers. In particular, tidal flaring rates would be enhanced at a
  stage when the SMBHs are still isolated in the cores of merging
  galaxies because of the mutual perturbation between merging
  galaxies, much earlier than the formation of SMBHBs. 
 \citet{roos81}  pioneered the discussions on the 
   stellar tidal disruptions by assuming that {merging} galaxies harbor
Sgr A*-like SMBHs and by taking into account perturbations by companion
 galaxies. However, it is unclear by how much the tidal
 disruption rates can be boosted in physical galaxy models
combining the correlations of central SMBHs and galactic bulges, 
how many tidal flares in the universe are contributed by
this merger phase, and how they would affect the constraint on the
merger history of SMBHs. As a first step toward addressing these
issues, in this paper we calculate the stellar-disruption rates during
galaxy mergers and investigate the prospect of using tidal flares to
probe multiple SMBHs in merging galaxies.

The outline of the paper is as follows. In \S~\ref{sec:lc}, we
introduce the basic loss-cone theory and the stellar-disruption
process in single SMBH systems. In \S~\ref{sec:lcmerge}, we describe
the stellar relaxation process in merging systems and generalize the
loss-cone theory to calculate the corresponding stellar-disruption
rate. We also discuss our results for different merger
parameters. Based on the calculated rates, we
investigate the contribution of tidal fares by merging galaxies
in \S~\ref{sec:num} and discuss our results and their
implications in \S~\ref{sec:dis}. 

\section{Loss-cone feeding in single SMBH system}\label{sec:lc}

We first calculate stellar-disruption rates for isolated galaxies
with single SMBHs, to prepare the basics for more complicated
calculations for merging galaxies. A star with mass $m_*$ and radius
$r_*$ would be tidally disrupted when it passes by an SMBH as close as
the tidal radius 
\begin{eqnarray}\label{rt}
  r_{t}&\simeq&r_*\left(\frac{M_\bullet}{m_*}\right)^{1/3}\\ 
  &\simeq& 4.9\times10^{-6} ~M_7^{1/3} \times \nonumber\\
  &&  \left(\frac{r_*}{\rsun}\right)
  \left(\frac{m_*}{\msun}\right)^{-1/3} ~{\rm pc} 
\end{eqnarray}
\citep{hills75,rees88}, where $\mb$ is the black hole mass,
$M_7=\mb/10^7~\msun$, and $\rsun$ and $\msun$ are, respectively, the
solar radius and mass. In the following, we assume $r_*=\rsun$ and
$m_*=\msun$ unless mentioned otherwise. For these solar-type stars,
when $\mb \ll 4\times10^7~\msun$, tidal disruption happens outside the
marginally bound orbit of the black hole, and collisions
between the bound stellar debris, as well as the subsequent accretion
on to the black hole, could produce an EM flare known
as the ``tidal flare'' \citep{rees88}. The criterion for stellar
disruption is then $J\le\jtd\simeq(2G\mb r_t)^{1/2}$, where $G$ is
Newtonian gravitational constant, $J$ is the specific angular
momentum, and $J_{\rm td}$ is the specific angular momentum
  corresponding to a pericenter distance of $r_t$. Here, the latter
approximation accounts for the fact that most stars are
disrupted along parabolic orbits, i.e., their specific binding energy
$\E\ll G\mb/r_t$. When 
$\mb>4\times10^7~\msun$, the marginally bound orbit 
of Schwarzschild black hole becomes greater
than the tidal radius, then the criterion for stellar depletion becomes
$J<\jmb$, where $\jmb$ denotes the specific angular momentum for
marginally bound geodesic. In general, $\jmb$ is a function of
black-hole spin and inclination relative to the equatorial plane, but
for simplicity we adopt the orientation-averaged value $\jmb=4G\mb/c$
\citep{kesden11}  in the following calculation, where $c$ is the speed
of light. For even greater black-hole mass
$\mb\ga10^9~\msun$, tidal disruption occurs inside the event horizon
of the central SMBH even when the black hole is {maximally}
spinning, so no tidal flare could be produced by disrupting solar-type stars
\citep{ivanov06,kesden11}.

As a result of tidal disruption and direct capture, a small fraction
of stars are lost from the system during their pericenter passages. In
a spherical system, the disruption rate of stars from distance
{$r$ to $r+ dr$} from the central SMBH is approximately 
\ba
   {d\Gamma}&\simeq&{4\pi r^2 dr\rho(r)\over m_*}{\theta^2(r)\over
     t_d(r)}
   \label{eqn:ratelc} 
\ea
\citep{frank76,syer99}, where $\rho(r)$ is the stellar mass density at
$r$, $t_d(r)$ is the dynamical timescale, and $\theta^2(r)$ estimates
the fraction of stars {subjected} to lose from the system. {The loss
fraction $\theta^2$ is dimensionless and can be interpreted
geometrically as a solid angle}, because at
$r$ the lost stars have velocity vectors pointing toward the SMBH
within an angle of $\thelc(r)=\jlc/J_c(r)$ and in an isotropic system
their fraction is $\theta^2=\thelc^2$. Here $J_c$ denotes the angular
momentum for circular orbit and is of order $r\sigma(r)$ given the
stellar velocity dispersion $\sigma(r)$. The cone-like region with
half-opening angle $\thelc$ toward the SMBH is therefore called ``loss
cone''. The isotropy of stellar distribution breaks down 
at the edge of loss cone when the
orbital-averaged rms velocity deflection angle $\theta_d(r)$ is much
smaller than $\thelc$ \citep{lightman77,cohn78}. Taking this effect
into account, careful analysis of the loss-cone structure suggests
that  
\ba
\theta^2=\min(\thelc^2,\theta_d^2/\ln\thelc^{-1})\label{eqn:theta2}
\ea
\citep{young77}. {Therefore}, when
$\theta_d\gg\thelc$ (``pinhole regime''), Equation~(\ref{eqn:theta2})
recovers $\theta^2=\thelc^2$, because the stars act as if the loss
cone does not exist and the system remains isotropic. On the other
hand when $\theta_d\ll\thelc$ (``diffusive regime''), the loss cone
becomes empty within one dynamical timescale, so afterwards only a
fraction $\theta_d^2/|\ln\thelc|$ of stars residing at the boundary
layer $\thelc\sim\thelc+\theta_d$ of the loss cone will be depleted
during one $t_d$. The total stellar disruption rate $\Gamma$ is
an integration of Equation~(\ref{eqn:ratelc}) over both pinhole and
diffusive regimes. 

To calculate $\rho(r)$, $t_d(r)$, and $\theta^2(r)$, a physical model
describing the stellar distribution in the host galaxy needs to be
specified. We consider only the bulge component of a galaxy because it
is the major source for stellar disruption. We model a bulge with a
spherical model with double power laws, i.e.,  
\begin{eqnarray}
\rho(r)&=&
\left\{
\begin{array}{lll}
\rho_b(r/r_b)^{-\gamma}\,\,\,\,\,\,\,\,(r\le r_b)\\
\rho_b(r/r_b)^{-\beta}\,\,\,\,\,\,\,\,(r_b<r<r_{\rm max})\\
0\,\,\,\,\,\,\,\,\,\,\,\,\,\,\,\,\,\,\,\,\,\,\,\,\,\,\,\,\,\,\,\,\,(r\ga \rmax)
\end{array},
\right.
\end{eqnarray}
where $r_b$ is the break radius, $\rho_b$ is the stellar mass density
at $r_b$, $\gamma$ and $\beta$ are, respectively, the inner and outer
power-law indices, and $\rmax$ is the cut off radius to prevent
divergence of the total stellar mass. The five model parameters,
$(r_b,\rmax,\rho_b,\gamma,\beta)$, are determined by the following
five physically motivated conditions 
\begin{itemize}
\item[1.] We define $r_b$ as the influence radius of SMBH\footnote{It
 is suggested that when $(\beta-\gamma)\ga1$, a practical definition
 for $r_b$ is that the mass deficit inside $r_b$ is $2M_\bullet$
 \citep{merritt06b,stone11b}, but the resulting $\rho_b$ differs from
 our fiducial value by only a factor of $(\beta-\gamma)/(3-\beta)$.} 
 such that the enclosed stellar mass is $2\mb$.

\item[2. and 3.] The values of $\gamma$ and $\beta$ are adopted from
  empirical galaxy models \citep{faber97,lauer05} and will be
  specified explicitly in the following calculations. In our fiducial
  model, $\gamma=1.75$ and $\beta=2$, so that the galaxy has an inner
  Bahcall-Wolf and outer isothermal profile \citep{bahcall76}. By
  varying $\gamma$ and $\beta$ ($\gamma,\beta<3$), our simplified
  galaxy model could reconcile with a variety of real galaxies. 

\item[4.] The total stellar mass enclosed in the radius $\rmax$ is
  $A\mb$, where $A=400$ so that  the SMBH-to-galaxy mass ratio
  satisfies the empirical correlation in the local university
  \citep[e.g.,][]{marconi03}. 

\item[5.] At the effective radius $r_e$, where the two-dimensional 
(2-D) surface-density
  isophote encloses half of the total galaxy mass, the stellar
  velocity  dispersion $\sigma_e$ satisfies the empirical correlation
  $M_\bullet\simeq10^8~(\sigma_e/200~\kms)^4~\msun$
  \citep{tremaine02}. Note that the stellar mass enclosed by the 3-D
  sphere of radius $r_e$ is $M_*(r_e)\simeq(0.36,0.32)A\mb$ when
  $\beta=(2,1.5)$, smaller than half of the galaxy mass. 

\end{itemize}

According to Jeans's equation in the isotropic limit
\begin{eqnarray}\label{eqn:jeans}
\frac{d(\rho\sigma^2)}{dr}+\frac{G\rho [M_*(r)+\mb]}{r^2}=0,
\end{eqnarray}
the velocity dispersion $\sigma\propto r^{-1/2}$ when $r\ll r_b$ and
$\sigma\propto r^{1-\beta/2}$ when $r\gg r_b$; therefore, we calculate
$\sigma$ with 
\begin{eqnarray}\label{eqn:sig}
\sigma(r)&=&
\left\{
\begin{array}{ll}
\sigma_b(r/r_b)^{-1/2}\,\,\,\,\,\,\,\,\,\,\,(r\le r_b)\\
\sigma_b(r/r_b)^{1-\beta/2}\,\,\,\,\,\,\,\,(r>r_b)\\
\end{array},
\right.
\end{eqnarray}
where $\sigma_b$ is the velocity dispersion at $r_b$. By applying
Equations~(\ref{eqn:jeans}) and (\ref{eqn:sig}) at $r_e$, we first derive 
\ba\label{eqn:re}
r_e=\frac{A_e+1}{2\beta-2}\frac{G\mb}{\sigma_e^2},
\ea
where $A_e\equiv M_*(r_e)/\mb$ and $\sigma_e$ is computed with
  $\mb-\sigma_e^4$ relation.  
 Then the model parameters $(r_b,\rmax,\rho_b)$ are calibrated
according to their definitions, and the results are $r_b\simeq
r_e[(6-2\gamma)/(3A_e-\beta A_e)]^{1/(3-\beta)}$, $\rmax\simeq
(A/A_e)^{1/(3-\beta)}r_e$, and $\rho_b=(3-\gamma)\mb/(2\pi
r_b^3)$. For example, our fiducial galaxy model with $\mb=10^7~\msun$,
$\gamma=1.75$, and $\beta=2$ corresponds to $r_b\simeq4.5$ pc,
$r_e\simeq260$ pc, and $\rmax\simeq820$ pc. 

Having specified the galaxy model, we now calculate the deflection
angle $\theta_d$ which determines $\theta^2$ in
Equation~(\ref{eqn:theta2}). Two-body scattering is an inherent
relaxation mechanism in stellar system and it gives a lower limit of
$\theta_2=J_2/J_c$ to $\theta_d$, where $J_2$ is the cumulative change
of $J$ due to two-body scattering during one dynamical
timescale. Because successive two-body scatterings are uncorrelated
(incoherent), we have $J_2=(t_d/t_r)^{1/2}J_c$, where   
\begin{eqnarray}
t_{\rm r}(r)&=&\frac{\sqrt{2}\sigma^3(r)}{\pi
  G^2m_*\rho(r)\ln\Lambda}\\ 
&=&\frac{2\sqrt{2}B^2}{(3-\gamma)\ln\Lambda} \frac{M_\bullet}{m_*}
  \times \nonumber\\
&&\left(\frac{\sigma}{\sigma_b}\right)^3\left(\frac{\rho}{\rho_b}\right)^{-1}
\frac{r_b}{\sigma_b} 
\end{eqnarray}
is the two-body relaxation timescale, $\ln\Lambda$ is the Coulomb
logarithm (we assumed a fiducial value of 5), and  
\ba\label{eqn:b}
B&\equiv&\frac{r_b}{G\mb/\sigma_b^2}\simeq\frac{3-\gamma}{(3-\beta)(\beta-1)}
\ea
is a correction factor of order unity. When two-body scattering
dominates the relaxation process, $J_2$ is an increasing function of
$r$, with the transition between pinhole and diffusive regimes
($J_2\sim\jlc$) being situated at $r\sim r_b$. The differential loss
rate $d\Gamma/dr$ (eq.~(\ref{eqn:ratelc})) scales as
$r^{9/2-2\gamma}$ in the diffusive regime ($r\ll r_b$) and as
$r^{-1-\beta/2}$ in the pinhole one ($r\gg r_b$); therefore, the
stellar disruption rate peaks at the transition regime at $r\sim
r_b$.  

Take our fiducial model with $\mb=10^7~\msun$, $\gamma=1.75$, and
$\beta=2$ for example. The critical radius where $\theta_2^2=\theta_{\rm lc}^2$ is
$\rcri\simeq2.3r_b$, and the total disruption rate due to
two-body relaxation is $\Gamma\simeq2.3\times10^{-5}~\peryr$,
consistent with previous calculations
\citep[e.g.,][]{magorrian99,syer99,wang04,brockamp11}. {If $\mb$
  increases, $\rcri/r_b$ will also increase, given the fact that
  $\theta_{\rm lc}^2$ is a decreasing function of $r/r_b$, and that
  $\theta_2^2\propto \mb^{-1}$ and $\theta_{\rm lc}^2\propto
  \mb^{1/3}$ at any $r/r_b$. On the other hand, the integrated
  stellar-disruption rate will decrease, mainly because the diffusive
  regime of loss cone becomes larger.} 
A more accurate calculation of $\Gamma$ could be
carried out by solving the diffusion equation in the 2-D
$\E-J$ space \citep[e.g.,][]{lightman77,cohn78,magorrian99,wang04}, but
it is considerably time-consuming and out of the scope of this
paper. Nevertheless, the present scheme gives good approximation to
the two-body disruption rate, and is sufficient 
to provide references for the sake of investigating the effects of
galaxy mergers on the stellar-disruption rate.  

\section{Enhanced loss-cone feeding during galaxy merger}\label{sec:lcmerge}

Because the loss cone is already ``full'' in the pinhole regime,
enhancing relaxation efficiency in this regime does not increase the
fraction of loss-cone stars, therefore would not increase
stellar-disruption rate. On the other hand, the loss cone in the 
diffusive regime is largely empty, so the disruption rate can
be enhanced if stellar relaxation in this regime becomes more
efficient. Enhancement of stellar relaxation in the
diffusive regime can be achieved by galaxy merger 
due to at least two processes. First, perturbation by
the companion galaxy would secularly change the stellar angular
momenta \citep{roos81}. Second, the triaxial gravitational potential
built up during  merger \citep{preto11,khan11} would drive stars to
galaxy center in a chaotic manner \citep{poon01}. In this section we
calculate the stellar-disruption rates due to the above two
processes, and we show the rate for each of the two SMBHs in the
merging system.  

\subsection{Basic Theory}\label{sec:theory}

{A companion galaxy would tidally torque the stellar orbits in the central galaxy, secularly changing the orbital elements. Given mass $M_p$ of the perturber and its distance $d$ from central galaxy, one can derive $GM_pr/d^3$ for the tidal force exerted by $M_p$ across a stellar orbit of radius $r\ll d$ in the central galaxy. The corresponding tidal torque on the stellar orbit is of magnitude $T_p\sim
GM_pr^2/d^3$.}  Because of the tidal torque, the
angular momentum of star changes coherently, i.e., $\Delta J\propto t$,
up to a timescale $t_\omega$, where $t_\omega$ is determined by the
shorter one between the dynamical timescale of the perturber and the
apsidal precession timescale of the stellar {orbit} \citep{bt08}. For
$t>t_\omega$, the torque on stellar orbit adds up stochastically and
in this case $\Delta J^2\propto t$. Therefore, averaged over a
timescale much longer than $t_\omega$, the tidal torque changes $J^2$
by an amount of $J_p^2=T_p^2t_\omega t_d(r)$ during each stellar
dynamical timescale. As a result, the deflection angle $\theta_d^2$ in
Equation~(\ref{eqn:theta2}) increases by an amount of
$\theta_p^2=(J_p/J_c)^2$. We note that the calculation of $J_p$ is
analogous to the calculation of angular-momentum change due to
resonant relaxation where the resonance torque is induced by the
grainy gravitational potential \citep{rau96,hop06}. 

Galaxy merger also increases the triaxiality of the gravitational
potential \citep{preto11,khan11}. \citet{poon01} showed that when the
triaxiality is large, a consistent fraction of stars are fed to the
loss cone in a chaotic manner and the loss cone remains full. Suppose
$f_c$ is the fraction of stars on chaotic orbits, the extra
contribution to stellar-disruption rate can be calculated by
replacing $\theta_d^2$ with $\theta_c^2=f_c\thelc^2\ln\thelc$
\citep{merritt04}. It has been shown that $f_c$ approaches unity when
the triaxiality becomes greater than $0.25$, but will rapidly decrease
to $0$ inside the influence radius of the central SMBH where the
gravitational potential is largely spherical \citep{poon04}. 

Because of tidal perturbation and triaxiality during galaxy merger,
the effective deflection angle $\theta_d^2$ increases to  
\ba
\theta_d^2=\theta_2^2+\theta_p^2+\theta_c^2,\label{eqn:thed2}
\ea
and  in the diffusive regime the loss-cone-limited 
deflection angle ($\theta^2$ in Equation~(\ref{eqn:theta2})) also
becomes larger.  Consequently, an
enhancement of stellar-disruption rate is anticipated. Now we have
prepared Equations~(\ref{eqn:ratelc}), (\ref{eqn:theta2}), and
(\ref{eqn:thed2}) to calculate the stellar-disruption rate in
merging galaxies. However, the equations are valid only in the
adiabatic approximation, i.e., the gravitational potential varies on a
timescale much longer than the typical timescale for stellar orbital
evolution. If the adiabatic condition is violated, the galaxy core
will be subject to significant heating and expansion on the dynamical
timescale \citep{ost72,mer01,boy07}. For the stars at $r\sim r_b$
which predominate the  loss-rate enhancement, the maximum timescale of
coherent angular-momentum change, $t_\omega$, is limited 
by the apsidal precession timescale, which is of order $t_d(r_b)$. The
timescale for chaotic orbital evolution is also of order
$t_d(r_b)$. The adiabatic limit therefore requires the 
orbital period of the merging galaxies
to be longer than $t_d(r_b)$. For
this reason, the following calculations are restricted to  $d>2 r_b$. 

\subsection{Stellar-disruption Rates}\label{sec:rate}

\begin{figure}
\plotone{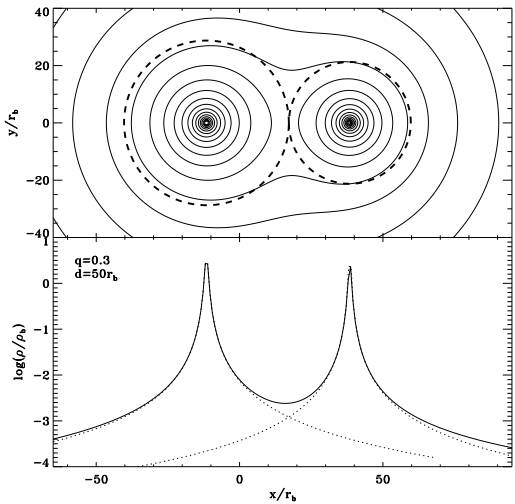}
   \caption{Upper: density contour in the mid-plane of a merging
   system with $M_\bullet=10^7~\msun$ and $q=0.3$. The two galaxies,
   both have $\gamma=1.75$ and $\beta=2$, are separated by $50r_b$
   where $r_b$ refers to the break radius of the main (bigger)
   galaxy. The dashed circles mark the tidal truncation radii.
   Lower: density distribution along the line connecting the two
   black holes (solid curve). The dotted lines show to the initially
   unperturbed density distributions.} 
   \label{fig:dens}
\end{figure}

We now calculate the stellar disruption rate for both galaxies
  in a merger. The black-hole and bulge 
components are modeled with the parameters $(\mb,\gamma,\beta)$, as is
described in \S~\ref{sec:lc}. The mass ratio of the galaxies, by
construction, equals the mass ratio of the SMBHs,
$q\equiv\mbs/\mbm\le1$, where the subscript $m$ denotes the quantity
for the bigger {\it main} galaxy and $s$ for the smaller {\it
  satellite} galaxy. As we have shown that the contribution to
stellar tidal disruptions is
dominated by the stars at the break radius of galaxy, 
we can approximately construct a merger
system of galaxies without loss of generality as follows.
Given the distance $d$ between the two
galaxy centers, the total stellar 
density at any location is approximated by summing the densities of
the two unperturbed  bulges. In this density field, each galaxy
approximately preserves its initial structure out to a radius
$\min(\rmax,\rtr)$, where $\rtr$ is the truncation radius due to
mutual tidal interaction, defined by the condition that the mean
densities within $\rtr$ are the same for the two truncated
galaxies. Figure~\ref{fig:dens} shows the density contours (upper
panel), as well as the density distribution along the line connecting
the two black holes (lower panel), for a merging system with
$\mb=10^7~\msun$, $q=0.3$, and $d=50r_b$. 

Given the configuration of the merging system, we calculated
$\theta_2$ and $\Gamma$ due to two-body relaxation for each of the two
galaxies. To calculate $\theta_p$ and $\Gamma$ due to tidal
perturbation, the perturber mass $M_p$ is derived by integrating the
stellar and black-hole masses in the perturber galaxy enclosed by
$\rtr$. Note that the perturber is the satellite galaxy when
calculating $\Gamma$ for the main galaxy, but can also be the main
galaxy when calculating $\Gamma$ for the satellite. To calculate
$\theta_c$ and $\Gamma$ due to chaotic loss-cone feeding, the
triaxiality of galaxy needs to be determined.  
But our model is axisymmetric by construction, so the triaxiality
cannot be derived self-consistently. We circumvent this inconsistency
by assuming that at any radius where the density increment induced by
the perturber excesses $\delta=20\%$ of  the initially unperturbed
density, a fraction of $f_c=50\%$ of stellar orbits are chaotic.
Otherwise $f_c=0$ if $\delta<20\%$.
The radial range where $f_c=50\%$ is insensitive to the choice of
$\delta$ because of the steep density profiles we adopted in the
following calculations. 

\begin{figure}
\plotone{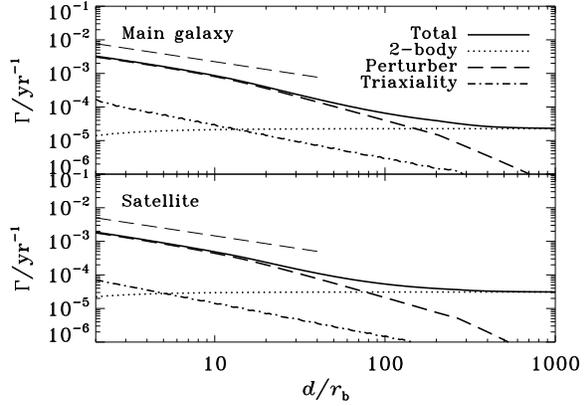}
   \caption{Stellar disruption rates as a function of galaxy
   separation for main (upper) and satellite (lower)
   galaxies. The dotted, dashed, and dash-dotted
   lines refer to {rates induced} by, respectively, two-body
   relaxation, tidal perturbation, and triaxial gravitational
   potential. {The thin dashed lines show the analytical solution
   $\Gamma\propto d^{-5/7}$} {in arbitrary units} {derived
   in Section~\ref{sec:para}.} The 
   model parameters are $(M_7,q,\gamma,\beta)=(1,0.3,1.75,2)$ and
   $r_b$ refers to the break radius of the main galaxy.} 
   \label{fig:rate}
\end{figure}

Figure~\ref{fig:rate} shows the stellar disruption rates as a
function of $d$ for both main (upper panel) and satellite (lower
panel) galaxies. The parameters are
  $(M_7,q,\gamma,\beta)=(1,0.3,1.75,2)$ by default. When $d \gg
100r_b \approx 450\, {\rm pc}$, the loss-cone 
filling in both galaxies is 
dominated by two-body relaxation (dotted lines) and the disruption rate
is identical to that for isolated single SMBH. As the distance shrinks
to $d\sim100r_b$, about $2r_e$ of the central galaxy, the
disruption rates induced by companion galaxies start
to exceed those due to two-body relaxation. This is because $\theta_p(r_b)$ becomes
greater than $\theta_2(r_b)$. As $d$ further decreases to $d \la
  10r_b \approx 45 \, {\rm pc}$,
$\theta_c(r_b)$ becomes greater than $\theta_2(r_b)$, so the
contribution to $\Gamma$ due to triaxial potential starts to exceed
that due to two-body relaxation. When the two galaxy cores are as
close as the break radius of the main galaxy, $\Gamma$ in
both galaxies have been enhanced by two orders of magnitude. In the
subsequent evolution with $d\la 2 r_b$ for which our simple scheme
  cannot be applied, the three-body interactions between
the two gravitationally bound SMBHs and the surrounding stars are
  expected to play an important role and to further enhance the
disruption rates \citep{ivanov05,chen09,chen11,wegg11}. 

{In a real merger, because galaxy orbitals are eccentric
  \citep{jiang08}, the distance $d$ will not decrease monotonically,
  but oscillate between the apocenter distance $r_{\rm apo}$ and
  pericenter distance $r_{\rm per}$, both distances decreasing with
  time due to dynamical friction. In this case, one can average the
  stellar-disruption rate over one orbital period according to
  $\bar{\Gamma}=\int_{r_{\rm per}}^{r_{\rm
      apo}}\Gamma(r)v_r^{-1}dr/\int_{r_{\rm per}}^{r_{\rm apo}}
  v_r^{-1}dr$, where $v_r$ denotes the radial velocity of galaxy at
  distance $r$. Since in our model, where $r/v_r\propto r^{\beta/2}$
  and $\Gamma(r)\propto r^{-\eta}$, both $\beta$ and $\eta$ are of
  order unity (see Section~\ref{sec:para}), we find that
  $\bar{\Gamma}$ differs from $\Gamma(d)$ by a factor of also order
  unity if we define $d\equiv(r_{\rm per}+r_{\rm apo})/2$. In this
  sense, the rates in Figure~\ref{fig:rate} can be used as the
  orbital-averaged stellar-disruption rates for galaxy mergers with
  eccentric orbits. We also note that} we may have underestimated the
contribution from triaxial potential, because 
in our model by construction $f_c$ vanishes inside about the influence radius
of black hole, as the stellar-density variation $\delta$ inside the
sphere of radius $r_b$ is small (e.g. lower panel of
Figure~\ref{fig:dens}). In real galaxies, however, chaotic orbits may
partially {exist} inside the influence radius of black hole
\citep{poon01}.

\subsection{Dependence of disruption rate on model parameters}\label{sec:para}

In \S~\ref{sec:rate}, we have shown that tidal perturbation by the
companion galaxy dominates the enhancement of $\Gamma$ in a
merger. The enhancement occurs when $\theta_p$ in the diffusive regime
exceeds $\theta_2$. As a result, the critical radius $r_{{\rm cri},p}$
that separates the pinhole and diffusive regimes is now determined by
$\theta_p(r_{{\rm cri},p})=\thelc$, and enhancement of
stellar-disruption rate requires that $r_{{\rm cri},p}<r_{\rm
  cri}$. Now we investigate in what mergers the condition $r_{{\rm
    cri},p}<r_{\rm cri}$ would be satisfied. 

According to $J_p(r_{\rm cri,p})=\jlc$ and the relation
\ba
   {M_\bullet\over M_p}&\propto& q^{3(1-\beta_p)/(2\beta_p)}
   \left(C_p\over C\right)^{3(3-\beta_p)/\beta_p} \times \nonumber \\
   &&\left(d\over r_b\right)^{\beta(\beta_p-3)/\beta_p},
\ea
where
\ba
C\equiv\frac{r_b}{G\mb/\sigma_e^2}\simeq B\left[\frac{A_e(3-\beta)}{6-2\gamma}\right]^{(2-\beta)/(3-\beta)},
\ea
we first derive the following scaling relation in the limit $r_{\rm
  cri,p}\la r_b$ and $\jlc=\jtd$ for the central galaxy:
\ba
&&{r_{\rm cri,p}\over r_b}\propto
B^{1/7}C^{-1/7}\mb^{-1/42}q^{3(1-\beta_p)/(7\beta_p)} \times \nonumber\\ 
&&\left(C_p\over C\right)^{6(3-\beta_p)/(7\beta_p)}
\left(d\over r_b\right)^{2\beta(\beta_p-3)/(7\beta_p)+6/7}.\label{eqn:rc2rb}
\ea
Since $1<\beta<3$ for the majority of galaxies \citep{lauer05}, Equation~(\ref{eqn:rc2rb}) suggests that in general enhancement of $\Gamma$ would occur when the perturbing galaxy is larger or the galaxy distance is smaller. When tidal perturbation dominates the loss-cone filling, according to Equation~(\ref{eqn:ratelc}) and $\mb\propto \sigma_e^4$, the rate $\Gamma$ in the limit $r_{\rm cri,p}\la r_b$ scales as
\ba
\Gamma\propto C^{-5/2}B^{-1/2}\mb^{7/12}(r_{\rm cri,p}/r_b)^{1/2-\gamma}.\label{eqn:gamm}
\ea
For our fiducial model with $\gamma=7/4$ and $\beta=2$, we can derive
$\Gamma\propto(d/r_b)^{-5/7}$, which is consistent with the
  numerical results given by the dashed lines at $d<{30}r_b$ in
  Figure~\ref{fig:rate}.  

\begin{figure}
\plotone{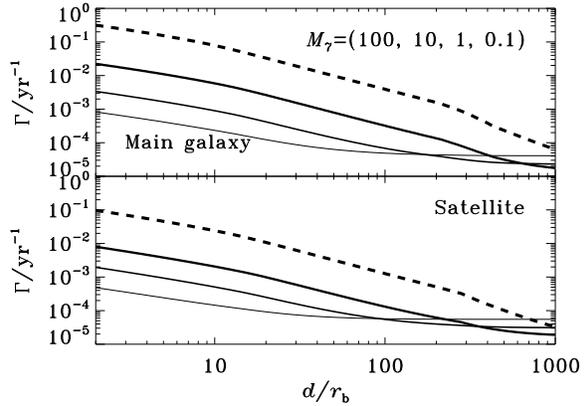}
   \caption{Total stellar disruption rates as a function of
   galaxy separation for different black hole masses. The model
   parameters are the same as in Figure~\ref{fig:rate}, except that
   $M_7=(100,10,1,0.1)$ from top to bottom with decreasing line
   thickness. The dashed lines indicate that central SMBHs are more
   massive than $10^8~\msun$ {and stars fall into the SMBHs
   without tidal disruption.}}
   \label{fig:rm}
\end{figure}

Figure~\ref{fig:rm} shows the dependence of $\Gamma$ on black hole
mass when $q=0.3$. Equations~(\ref{eqn:rc2rb}) and (\ref{eqn:gamm})
suggest that $\Gamma\propto M_\bullet^{(24+\gamma)/42}$ when $q$ and
$d/r_b$ are fixed. The enhanced stellar-disruption rates in
Figure~\ref{fig:rm} generally agree with this scaling when
$\mb\la4\times10^7~\msun$. When $\mb>4\times10^7~\msun$, the
dependence of $\Gamma$ on $\mb$ steepens because {direct capture of stars by SMBH (GR effect) becomes important, such that the scaling of loss-cone size changes from $\jlc^2\propto\mb^{4/3}$ to
$\jlc^2\propto\mb^{2}$}. When $\mb\ga10^8~\msun$, the
loss-cone stars will be directly captured by the central SMBH without
producing tidal flares if the SMBH is non-rotating or rotates
slowly \citep{ivanov06,kesden11}, and the corresponding curves are
shown in dashed lines. Note that even when the SMBH in the main
galaxy is more 
massive than $10^8~\msun$, the merging system could still produce
tidal flares, due to the existence of a smaller SMBH in the satellite
galaxy. We found that when $1<M_7 \la 10$, the
event rates of tidal flares can be as high as $\sim10^{-2}~\peryr$ as $d$
shrinks to about $r_b$.

\begin{figure}
\plotone{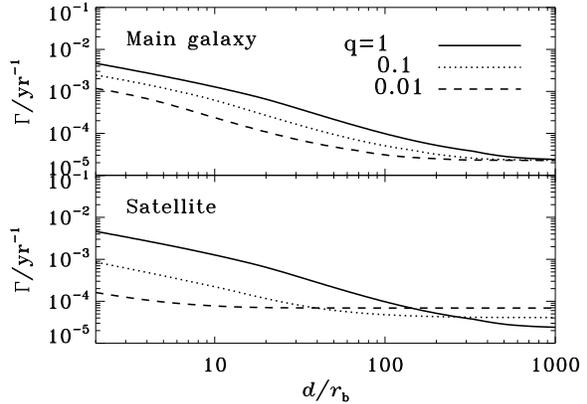}
   \caption{Stellar disruption rates as a function of galaxy
   separation for different $q$. The other parameters are the same as
   in Figure~\ref{fig:rate}.} 
   \label{fig:rq}
\end{figure}

Figure~\ref{fig:rq} shows the dependence of $\Gamma$ on the mass ratio
$q=\mbs/\mbm\le1$ of the two black holes, {while $\mbm$ is fixed}. For both main and satellite
galaxies, the enhancement of $\Gamma$ becomes more significant as $q$
increases. It is worth noting that even $q$ is as small as $0.01$,
the stellar disruption rate in the main galaxy can still be
enhanced by two orders of magnitude when $d$ shrinks to about
$r_b$. {We also find that the enhanced
stellar disruption rate in the satellite is more sensitive to
$q$ than that in the main galaxy. This is because the baseline stellar-disruption rate, i.e., the rate for single black hole in isolated galaxy, changes with $q$ for satellite galaxy, but does not vary for the main galaxy since  in the calculation $\mbm$ is fixed (e.g. see Equation~[\ref{eqn:gamm}]).} Quantitatively speaking, according
to Equations~(\ref{eqn:rc2rb}) and (\ref{eqn:gamm}), {when varying $q$ while keeping $\mbm$ fixed,} the enhanced
stellar disruption rate for the main galaxy scales as
$q^{3(1-\beta_s)(1-2\gamma)/(14\beta_s)}$, while the rate
for the satellite scales as
$q^{(24+\gamma_{{s}})/42-(1-2\gamma_{{s}})(3+\beta\beta_s-3\beta_{{s}})/(14{\beta})}$. {For example, given $(\gamma,\beta,\gamma_s,\beta_s)=(1.75,2,1.75,2)$,  one can derive $\Gamma\propto q^{15/56}$ for the main galaxy and $\Gamma\propto q^{59/84}$ for the satellite.}

\begin{figure}
\plotone{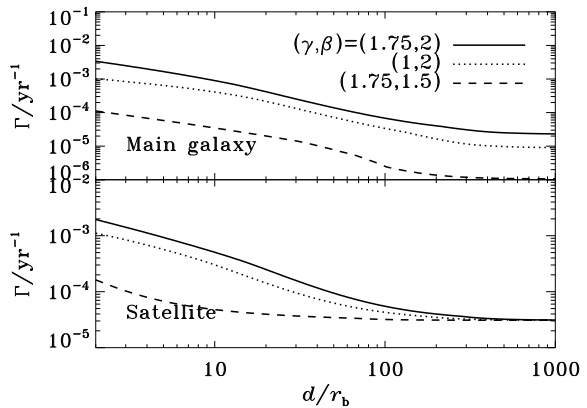}
   \caption{Stellar disruption rates as a function of galaxy
   separation for different density profiles in the main galaxy. The
   other parameters are the same as in Figure~\ref{fig:rate}.} 
   \label{fig:pro1}
\end{figure}

Figure~\ref{fig:pro1} shows the variation of stellar disruption
rate when the density profile of the main galaxy changes. For the main
galaxy, when the inner power-law index $\gamma$ decreases from $1.75$
to $1$, the stellar disruption rates due to two-body relaxation
and tidal perturbation both drop by a factor of a few, because of the
slight decrement of the stellar density at $r\sim r_b$. Meanwhile, the
dependence of $\Gamma$ on $d/r_b$ at $d\la10r_b$ changes from
$(d/r_b)^{-5/7}$ to $(d/r_b)^{-2/7}$, resulting in an
even smaller rate at $d\sim r_b$.  
When the outer power law index $\beta$ decreases
from $2$ to $1.5$, the stellar disruption rates in the main
galaxies drop approximately by  a factor of $20$. This is because the
galaxy with shallower outer density profile is more spatially extended
and has lower central density. For the satellite, when $\gamma$ or
$\beta$ of the main galaxy decrease, the enhancement of stellar
disruption rate occurs at smaller $d/r_b$ and becomes weaker for
a fixed $d/r_b$. This is because $r_b$ of the main galaxy becomes
greater as $\gamma$ or $\beta$ decrease, so that for the satellite the
physical distance of the perturber increases if $d/r_b$ is fixed. We
notice that when $\beta=1.5$, the stellar disruption rate in the main
galaxy remains lower than that in the satellite as $d$ decrease. This
result implies that in mergers where the main galaxies have low
  surface 
brightness, the tidal flares are mostly contributed by the satellite
galaxies. 

\begin{figure}
\plotone{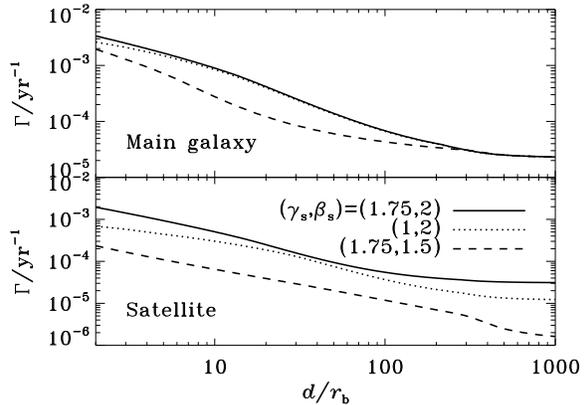}
   \caption{Same as Figure~\ref{fig:pro1}, but varying $(\gamma_s,\beta_s)$ in the satellite galaxy.}
   \label{fig:pro2}
\end{figure}

When the density profile of the satellite galaxy is varying, the
resulting stellar disruption rates are shown in
Figure~\ref{fig:pro2}. In general, the dependence of $\Gamma$ on the
density profile can be understood in the light of the analysis for
Figure~\ref{fig:pro1}, except that now the role between the
main and satellite galaxies switches. However, one difference is that
when $d$ shrinks to about $r_b$, the disruption rate in the main galaxy
is not sensitive to the density profile of 
the satellite. This is because when $d\sim r_b$ the stellar cusp
surrounding the SMBH in the satellite is almost completely striped off
by the tidal filed of the main galaxy, so for the main galaxy the
perturbing mass is approximately $M_{\bullet,s}$.  

{Figures~\ref{fig:rate}--\ref{fig:pro2} showed that galaxy merger
  starts enhancing stellar-disruption rate when the two galactic nuclei
  are still widely apart, well before the two SMBHs become
  gravitationally bound. The boost factor for each SMBH incorporated
  is about $10^2(\mb/10^7~\msun)(d/r_b)^{\mu}(q/0.3)^{\nu}$, where
  $\mb$ is the mass of the subject black hole, $r_b$ refers to the
  break radius of the more massive galaxy, and $\mu$ and $\nu$ are
  indices depending on the density profiles of the two galaxies. Less
  massive black holes have smaller boost factors because prior to
  merger they already have higher stellar-disruption rates. The exact
  boost factor depends on the stellar-disruption rate prior to galaxy
  merger, which deserves some discussion. When calculating $\Gamma$
  for isolated galaxies, we considered only two-body relaxation but
  not more efficient relaxation processes, such as resonant
  relaxation, perturbation by massive objects, or relaxation processes
  in triaxial gravitational potential
  \citep[e.g.][]{rau96,perets07,merritt04}. Resonant 
relaxation enhances stellar-disruption rate only mildly, less than a
factor of a few in typical galaxies \citep{rau98}. Massive perturbs,
such as molecular clouds and stellar-mass black holes, if highly
concentrate inside the influence radius of an SMBH, in principle could
enhance the stellar-disruption rate by orders of magnitude
\citep{perets07}. But such galactic nuclei could only be transient,
because large concentration of massive perturbs normally corresponds
to short relaxation timescale. On the other hand, weak triaxiality
seems intrinsic to galaxies, suggesting that chaotic loss-cone feeding
may be important prior to galaxy mergers. If we use Equation
  (119) in \citet{mer11}\footnote{{\citet{mer11} used
      $\gamma=1.5$ to derive the stellar-disruption rates. To derive
      rates for different $\gamma$, we used the scaling relation
      between $r_{\rm crit}$ and $\gamma$ above Equation~(116) in their
      paper.}} to estimate the stellar-disruption rate induced by
  triaxial potential inside the black-hole influence radius, meanwhile
  use formulae derived in Sections~\ref{sec:lc} and \ref{sec:theory}
  with $f_c=0.1$ to calculate the rate due to chaotic orbits outside
  the black-hole influence radius, then the total disruption rates for
  isolated fiducial galaxies become
  $\Gamma\simeq(5.8,8.9,47)\times10^{-5}~\peryr$ when
  $\mb=(10^6,10^7,10^8)~\msun$.} {For comparison, the rates due to
  two-body relaxation only are
  $(4.0,2.3,1.4)\times10^{-5}~\peryr$. The  difference is the greatest
  in the case of $\mb=10^8~\msun$, because the ``gap'' between $\rcri$
  and $r_b$ is the largest. These results suggest that only in the
  most massive galaxies with  $\mb\ga10^8~\msun$ could intrinsic
  triaxiality make the enhancement of stellar-disruption rate less
  significant.}  

\section{Contributions of tidal flares by merging galaxies}\label{sec:num}

In a synoptic sky survey, the probability of catching tidal flares
in merging galaxies
does not depend only on the stellar-disruption rate, but also on the 
duration of galaxy mergers. In other words, the fractions of tidal
flares in merging and in normal galaxies are proportional to the
  numbers of stellar-disruption events produced during,
    respectively, the merger and the quiescent phases. Since the 
duration of a galaxy merger is determined by the dynamical friction
timescale, $t_{\rm df}$, the fraction of tidal flares in merging
galaxies is proportional to the typical number of tidal stellar disruptions,
$n = t_{\rm df}\Gamma$. Given the distance $d$ between two merging
galaxies, we calculate the dynamical friction timescale as 
\ba
t_{\rm df}(d)=\left|\frac{d}{\dot{d}}\right|\simeq{M_g\over
  M_s}{t_d(d)\over\ln(M_g/M_s)}, 
\ea
\citep[{see eq.~[8.13] in}][]{bt08}, 
where $M_g(d)$ here refers to the stellar mass enclosed by the
  radius $d$ in the main galaxy and $M_s$ is the total mass of the 
truncated satellite. When the two galaxies are distant and $\Gamma$ is
not enhanced, the total number of disrupted stars is proportion to the
dynamical friction timescale, which is 
\ba
t_{\rm df}&\simeq& q^{-1}t_d(d)/\ln(M_g/M_s)\label{eqn:tdfn}
\ea
We refer to this early evolutionary stage as phase I. During phase I,
the dependence of $t_{\rm df}\Gamma$ on $d^{\beta/2}$ implies that the
majority of tidal flares are contributed by wide galaxy pairs. When loss-cone feeding is enhanced due to tidal perturbation by the
companion galaxy, the stellar-disruption rate $\Gamma$ increases with
decreasing $d$. We refer to this later evolutionary stage as phase II,
and we note that main and satellite galaxies enter phase II at
different times. During phase II, the dependence of $t_{\rm df}\Gamma$
on $d$ flattens compared to that in phase I, implying a enhanced
detection rate of tidal flares in close galaxy pairs.  

\begin{figure}
\plotone{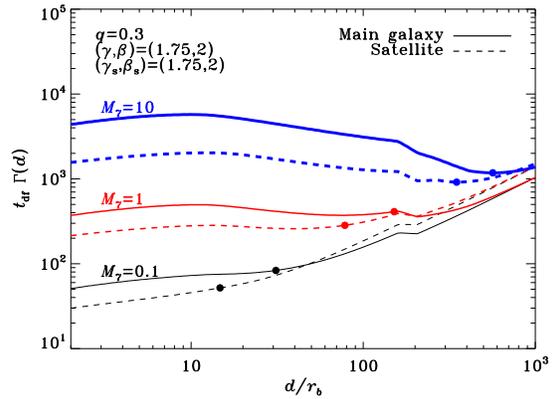}
   \caption{Typical number of disrupted stars contributed by
   main (solid) and 
   satellite (dashed) galaxies at different separations. Lines with
   decreasing thickness refer to systems with decreasing $M_7$. The
   other parameters are the same as in Figure~\ref{fig:rate}.} 
   \label{fig:sn}
\end{figure}

Figure~\ref{fig:sn} gives
the typical number of disrupted stars ($t_{\rm
  df}\Gamma$) as a function of $d$ in our fiducial model. In the
calculation, we did not consider the decrease of stellar density due
to tidal disruption, because the total mass of disrupted stars is
negligible with respect to the stellar mass in the initial
condition. In general, when the merger is in phase I, $t_{\rm
  df}\Gamma$ scales as $d$, as predicted. During this phase, $t_{\rm
  df}\Gamma$ is not sensitive to the total mass of the system
  as long as $q$ is fixed, because
massive systems where $\Gamma$ is larger have shorter $t_{\rm
  df}$. Note that before the main galaxy (solid curve) enters phase
II, the satellite galaxy (dashed curve) contributes comparable number
of, if not more, tidal flares. {This is because when two-body relaxation predominates the loss-cone filling, smaller galaxies have smaller diffusive loss cones, therefore will have higher stellar-disruption rates, as is explained in the end of Section~\ref{sec:lc}.}  When the galaxy mergers enter phase
II, which is marked by the dots, the $t_{\rm df}\Gamma$ curve
flattens, indicating an enhanced contribution of tidal flares by closer galaxy
pairs. During this phase, the contribution of tidal flares from main
galaxy is typically greater than that from satellite. 

\begin{figure*}
\plotone{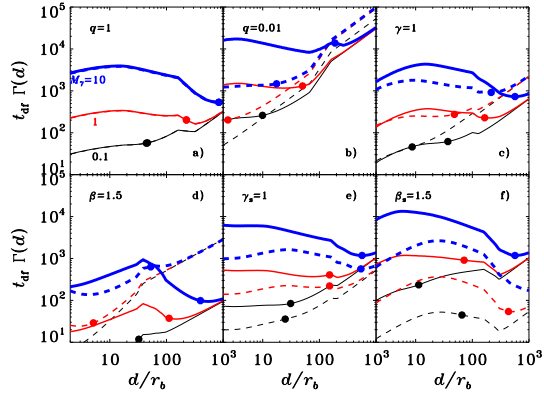}
   \caption{Same as Figure~\ref{fig:sn} but varying one model parameter, which is indicated at the upper-left corner of each panel.}
   \label{fig:num}
\end{figure*}

Figure~\ref{fig:num} shows the dependence of $t_{\rm df}\Gamma$ on
different parameters of galaxy merger, which are summarized 
as follows.
  
\begin{enumerate} 

\item Comparing panels (a) and (b), one can see 
  that during phase I, $t_{\rm
  df}\Gamma$ scales as $q^{-1}$, a characteristic relation due to
  dynamical friction. In phase II, as $q$ decreases from $1$ to
  $0.01$, $t_{\rm df}\Gamma$ for the main galaxy increases by a
  factor of 10 relative to that in the fiducial case in Figure
  7, but that for the satellite does not significantly change.  

\item When $\gamma$ of the main galaxies decreases from $1.75$ to $1$
  as shown in panel (c), in phase I the main galaxy contributes
  slightly less tidal flares compared to that in the
  fiducial model, because $\Gamma$ is smaller as $\gamma$
  decreases. Meanwhile, the satellite contributes slightly more tidal 
  flares because of longer $t_{\rm df}$. As a result, the relative
  contribution of tidal flares by the satellite becomes greater during
  phase I. During phase II,  $t_{\rm df}
  \Gamma$ for both main and satellite galaxies increases as $d$
  decreases from about $100 r_b$ to $10r_b$. As $d$ becomes smaller
  than $10r_b$, the typical number of stellar disruption decreases
  more steeply with smaller $d$ compared to that in the case with
  $\gamma=1.75$, because of less enhancement of stellar disruption
  rate as shown in Figure~\ref{fig:pro1}. 

\item Panel (d) shows that when $\beta$ of the main galaxy decreases
  from $2$ to $1.5$, the number of tidal flares from the
  main galaxy drops during both phases I and II by about one order of
  magnitude relative to that in the fiducial case, because $\Gamma$
  decreases significantly. While that from the satellite 
  increases in phase I due to longer $t_{\rm df}$ and significantly
  drops during phase II because of greater physical distance between
  the galaxies. As a result, tidal flares are dominantly from the
  satellite galaxies during phase I, and almost equally contributed by
  the main and satellite galaxies during phase II. 

\item Panel (e) indicates that when $\gamma_s$ of the satellite
  galaxy decreases from $1.75$ to $1$, the tidal flares contributed by
  the satellite become slightly less than those in the fiducial case
  in both phases I and II. During phase II, when $2 r_b \la d<10r_b$, the
  number of tidal flares contributed by the main galaxy slightly
  increases relative to that in the fiducial case, because the
  satellite is more severely tidally truncated so that $t_{\rm df}$
  becomes longer.   

\item When $\beta_s$ of the satellite galaxy decreases from $2$ to
  $1.5$ as shown in panel (f), the tidal flares contributed by
  the satellite 
  during phase I are one order of magnitude less than those in the
  fiducial case. During phase II, the main galaxy contributes more
  tidal flares than in the fiducial case, because now the satellite is
  more susceptible to tidal stripping and $t_{\rm df}$ is much longer
  than that in the fiducial case. 

\end{enumerate}

Figures~\ref{fig:sn} and \ref{fig:num} suggest that during
phase I the total number of disrupted stars scales roughly as  
\ba
n_{\rm I}(q,d)\sim10^3q^{-1}(d/10^3r_b)^{\beta/2},\label{eqn:nI}
\ea
insensitive to the total mass of the system or the stellar density
profiles of the merging galaxies. During phase II, when the curve
of $t_{\rm df}\Gamma$ is nearly independent of distance $d$, the total
number becomes about 
\ba
n_{\rm II}(q,M_7)\sim200~q^mM_7^n,
\ea
where $m\simeq3(1-\beta_s)(1-2\gamma)/(14\beta_s)-1$ and
$n=(45+\gamma)/42$ are power-law indices derived from
Equations~(\ref{eqn:rc2rb}), (\ref{eqn:gamm}) and (\ref{eqn:tdfn}). It
is worth noting that each merger investigated above involves only two
galaxies. However, mergers of group galaxies are also common 
  and the tidal disruption rates are expected
  to be even more heavily enhanced because of stronger perturbations
  and larger triaxiality. 

\section{Discussions}\label{sec:dis}

Formation of SMBHBs at galaxy centers is anticipated in the paradigm
of hierarchical galaxy formation \citep{beg80}, and coalescence of the
binaries is predicted to induce recoiling velocities on the
post-merger SMBHs \citep{cen10}. In our previous works
\citep{chen08,chen09,chen11,liu09}, we investigated the possibility of using
tidal-disruption flares to identify gravitationally bound SMBHBs of
sub-pc separations in galactic nuclei. Recently, off-nuclear tidal
flares have also been suggested in the literature to be probes of
recoiling SMBHs
\citep{komossa08,mer09,stone11a,stone11b,li12}. However, an off-center 
tidal flare can also be produced by SMBHs embedded in merging
galaxies. In this paper, we calculated the tidal flare rates produced
by dual SMBHs in a particular evolutionary stage when the two SMBHs
are still unbounded to each other and isolated in the cores of merging
galaxies. We considered three major processes responsible for
the loss-cone feeding in the merger system, namely, two-body
stellar relaxation, tidal perturbation by the companion galaxy,
and chaotic stellar orbits in triaxial gravitational potential.

By employing an analytical model to calculate the stellar
disruption rates for both SMBHs in the two merging galaxies, we
found that prior to the formation of SMBHB the stellar
disruption rate would be enhanced by as large as two orders of
magnitude in both galaxies. The enhancement is dominated by tidal
perturbation and occurs when the two galaxies are so close that the
stars inside the influence radius of the central SMBH are
significantly perturbed. We have shown that the enhanced stellar
disruption rate depends on the masses, mass ratio, and density
profiles of the two galaxies, as well as the distance $d$ between the
two galaxy cores. In the fiducial model with
$(M_7,q,\gamma,\beta)=(1,0.3,1.75,2)$, the enhancement starts when the
perturber galaxy approaches approximately twice the effective radius
of the central galaxy ($d\simeq 2r_e$). In more massive systems with
$M_7>10$, where the stellar disruption rates due to two-body
relaxation are generally lower, the enhancement starts as soon
as $d$ shrinks to $10r_e$. As a result, the phase with enhanced 
stellar-disruption rate extends to an  evolutionary stage much earlier
than the formation of bound SMBHB, which  considerably increases
the detection rate of wide SMBH pairs in tidal-flare surveys.  
When $d$ shrinks to about the influence radius of the central SMBH
($d\sim 2 r_b$), the stellar disruption rate in the fiducial
model increases to $3\times10^{-3}~\peryr$ in the main galaxy and to
$2\times10^{-3}~\peryr$ in the satellite. Compared to the peak rates
in the later evolutionary stages with gravitationally-bound binary SMBHs
\citep[e.g.][]{chen09,chen11,wegg11}, the total stellar-disruption rates
before SMBHs become bounded are smaller by only a factor of a
few. In more massive or equal-mass ($q>0.3$) mergers, the
stellar disruption rates could be even higher. 

The above results showed that the tidal disruption rates by off-center SMBH 
pairs in merging galaxies are several order magnitudes higher than
those by recoiling off-nuclear SMBHs \citep{komossa08,li12,stone11b},
implying that off-center tidal disruption flares would be overwhelmed
by the SMBH pairs in merging galaxies. Therefore, it would be
challenging to distinguish recoiling SMBHs in off-center tidal
disruption flares. One possible way to distinguish the two kinds of
off-center tidal disruptions may be to identify the evolutionary
stages of galaxies. Recoiling SMBHs are in galaxies at late stages of
mergers, while un-bounded SMBH pairs are in galaxies at early or
middle stages of mergers. Early stages of major mergers when
  galaxies are widely separated may be identified with the disturbed 
morphology of host galaxies. However, morphological signatures
  of galaxy merger are weak during the middle or late stages of major
  mergers, as well as during the whole stages of minor mergers,
therefore it would be also a challenge to identify these merger
stages. 
{Another difference may be among the properties of star clusters
  around the off-center SMBHs. A recoiling SMBH is expected to
reside in an ultra-compact bound star cluster of mass much
  smaller than the black-hole mass, of size much smaller than the
  black-hole influence radius, and of stellar-velocity
  dispersion much larger than that of host galaxy
  \citep{mer09,li12}. It may also associate with a massive cloud of unbound stars, whose mass is
  comparable to the black-hole mass, size comparable to the
  black-hole influence radius, and stellar-velocity dispersion comparable to or greater than that of the host galactic nuclei \citep{li12}. While
  the star clusters hosting the the secondary black holes in minor mergers are the remnants of the tidally truncated satellite galaxies. These clusters are orders of magnitude heavier than the
  secondary SMBHs, their sizes are much larger than the influence radii
  of the secondary or the primary SMBHs, and their stellar-velocity
  dispersions are comparable to those of typical
  dwarf galaxies but significantly smaller than those of the primary
  galactic nuclei. Therefore, the two types of star clusters should differ significantly in their sizes, stellar-velocity dispersions, and the mass ratios between SMBHs and star clusters, which could be
  identified with deep photometrical and spectroscopical observations.}

When a pair of SMBHs evolve to about the influence radius, $d \sim
  r_{\rm b}$, the enhanced stellar-disruption rates can be as high as 
$10^{-2}~\peryr$. For such a high tidal disruption rate, multiple
 tidal flares may occur in the same galaxy 
within a time span of decades. Unlike the recurring tidal flares in
binary or recoiling SMBH systems, the flares in merging galaxies are
contributed by wide SMBH pairs separated by $r_b\sim1-10$ pc
(depending on black hole mass and galaxy density profile). Note that
a separation of $10$ pc at redshift $z=0.1$ ($1$) corresponds to an
angular size of $5$ ($1$) milliarcsec (mas). As a result, spatial
offsets between successive tidal flares in such a merging system may
be detected by instruments such as {\it Gaia} and {\it
  LSST}\footnote{{See Section 3.7 of LSST Science Book
Version 2.0, http://www.lsst.org/lsst/scibook}}. Figures~\ref{fig:rm}
  and \ref{fig:rq} imply that such {\it 
  flip-flop} flares could occur in the galaxy mergers with
$10^7~\msun<\mb<10^8~\msun$ and $q>0.1$. The mergers with $\mb<10^7$
could not produce recurring flares because the stellar-disruption
rate is too low. When $10^7~\msun<\mb<10^8~\msun$ but $q<0.1$, most
flares are produced in the main galaxy; therefore, the recurring flares
are unlikely to display spatial offset, and would be indistinguishable
from those in binary or newly-formed recoiling SMBH systems. When
$\mb>10^8~\msun$, the SMBH in the main galaxy would directly capture
stars, mostly without producing flares, while the SMBH in the
satellite could still produce tidal flares if
$M_{\bullet,s}<10^8~\msun$. In the last case, although the recurring
flares occur at the same sky position, they should be displaced from
the center of the minor merger by an amount of $r_b\ga10$ pc. Sources
with such high flaring rate and large off-center displacement cannot
be produced by binary or recoiling SMBHs. The above discussions
  suggest that with the aid of telescopes with high spatial
  resolution, the cause of the recurring tidal flares can be
  distinguished.  

In the universe, the fraction of tidal flares contributed by galaxy
mergers is proportional to the total number of the disrupted stars
during merger.  During each merger, the
number of tidal flares contributed by phase I, when the separations of
galaxies are about $r_e\la d\la 10r_e$, is about $n_{\rm
  I}\sim10^3q^{-1}(d/10^3r_b)^{\beta/2}$ [Figures~(\ref{fig:sn}) and
  (\ref{fig:num})]. The scaling $n_{\rm I}\propto q^{-1}$ implies that
$n_{\rm I}$ is determined mainly by minor mergers. Suppose a
  galaxy experiences $N$ mergers during a Hubble time ($\sim10^{10}$
yr), then during one duty {cycle}, the number of tidal flares
contributed by the isolation phase is about $n_s\sim2\times10^5/N$, if
two-body scattering is the dominant relaxation process. Since a galaxy
with $M_7=1$ (10) at redshift $z=0$ has experienced typically $N\sim10$
(100) mergers and most mergers have $q\sim0.1$
\citep{hopkins10}, according to the ratio $n_{\rm I}:n_s$, we
  find that about $\sim5\%$ ($50\%$) tidal 
flares are contributed by phase-I galaxy mergers ($d\sim10^3r_b$). 

For typical mergers with $q\ll1$, according to Figure~\ref{fig:num},
the majority of the tidal flares are produced in satellite galaxies
during phase I, unless the satellite galaxies have low surface
brightness. This result implies 
that a large fraction of genuine tidal flares would be displaced by
several $r_e$ from the centers of the merging systems. Given that an
offset of $2r_e\sim 500$ pc corresponds to $250$ (60) mas at $z=0.1$
(1), these offset tidal flares could be misidentified as supernovae or
gamma-ray bursts by careless classification schemes. They may also 
be mistaken as ``{\it naked}'' recoiling quasars  \citep[e.g.,][]{komossa08} 
or ``{\it orphan transients}'' {\citep[X-/$\gamma$-ray transients either uncorrelated with bursts in low-energy bands or without detection of optical counterparts, e.g.,][]{horan09}} because of the relative dimness of
  the satellites. The mis-identification could be very common in
  massive galaxies, because the physical scale of $r_e$ is larger.

During phase II when the stellar disruption rates are
  enhanced by galaxy mergers, the main galaxies would
  contribution typically more than half of the tidal flares, unless
  $q\sim1$ or the surface brightnesses of the main galaxies are
  low. This result indicates that in an advanced merger, where the
  separation between the two galaxy cores is less than the effective
  radius of the main galaxy, the tidal flares preferentially reside in
  the massive nucleus of the system. According to Figures~\ref{fig:sn} and
  \ref{fig:num}, the number of tidal flares contributed by such
  advanced {merger} phase {does} not depend on $d/r_b$, and scales as
  $n_{\rm II}\sim200~q^mM_7^n$, where $m<0$ and $n>0$ are analytical
  indices derived in \S~\ref{sec:num}. Therefore, the biggest
  contribution is expected to come from minor mergers in massive
  systems. 
Because $n_{\rm II}$ is typically less than $n_{\rm I}$, the
contribution of tidal flares from phase II is typically smaller than
that from phase I. However, for the most massive systems with
$M_7\ga10$ in which the main SMBHs mostly swallow the stars
  without producing tidal flares, one major merger ($q>0.3$), or one
minor merger 
($q \la 0.3$) between galaxies of low surface brightness, would produce
more tidal flares in phase II than in phase I. In these particular
systems, a greater fraction of tidal flares would be contributed by
close SMBH pairs with separations $10\la d/r_b\la100$. 

It is important to know the relative contributions of tidal
flares by single ($n_s$), binary ($n_b$), and merging SMBH systems
($n_{\rm I}$ and $n_{\rm II}$). The total number of flares produced by
a recoiling black holes is typically smaller than $10^3$
\citep{komossa08,stone11a,li12}, therefore negligible in the
comparison.  According to \citet{chen11}, during the lifetime of an
SMBHB with $q\ll1$, the interaction between the binary and the
surrounding dense stellar cusp will produce a number of
$n_b\simeq7\times10^4~q^{(2-\gamma)/(6-2\gamma)}M_7^{11/12}$ of tidal
flares. Suppose a galaxy on average experiences $N$ mergers and $M$ ($M\le N$) of them 
result in the formation of SMBHBs. Then being averaged by one duty
{cycle} of galaxy merger, $n_s:n_b:(n_{\rm I}+n_{\rm II})$ is about
$20:5M:N$, where {we used $q=0.1$ because minor mergers are the
  most common \citep[][]{vol03,hopkins10,mcw12b}. For galaxies of
  total masses $(10^9,10^{10},10^{11})~\msun$, typical $N$ are
  $(1,10,10^2)$ \citep{hopkins10} or} {significantly higher}
{\citep{mcw12a,mcw12b,bed13}, while $M$ are predicted to be
  greater 
  than $1$ \citep{vol03}. These numbers highlight} the significant
contribution of tidal flares from 
merging systems with multiple SMBHs. To give more accurate
calculations, one has to combine the cosmic merger history
of galaxies, as well 
as the formation rate of SMBHBs of different masses and mass
ratios. Such calculations and the assessment of their uncertainties
are beyond the scope of the current paper and will be addressed
  in a future paper. 

\acknowledgments

We are grateful to Shuo Li, Zuhui Fan, Rainer Spurzem, and Thijs
Kouwenhoven for helpful comments. We also thank Alberto Sesana for
earlier discussions on this topic. This work is supported by the
National Natural 
Science Foundation of China (NSFC11073002). F.K.L. also thanks the
support from the Research Fund for the Doctoral Program of Higher
Education (RFDP), and X.C. acknowledges the support from China Postdoc
Science Foundation (2011M500001).

\end{document}